\documentclass{elsart3p}

\usepackage{graphics}
 \usepackage{epsfig}

\usepackage{amssymb}

\begin{document}

\begin{frontmatter}



\title{Spectroscopic evidences of quantum critical charge fluctuations
in cuprates}


\author[a]{M. Grilli, S. Caprara, C. Di Castro, and D. Suppa}

\address[a]{SMC - Istituto Nazionale per la Fisica della Materia,
 and Dipartimento di Fisica, Universit\'a "La Sapienza",
 Roma, Italy}

\begin{abstract}
We calculate the optical conductivity in a clean system of quasiparticles coupled to
charge-ordering collective modes. The absorption induced by these
modes may produce an anomalous frequency and temperature dependence of 
low-energy optical absorption in some cuprates. However, the coupling with lattice 
degrees of freedom introduces a non-universal energy scale leading to scaling violation
in low-temperature optical conductivity.

\end{abstract}

\end{frontmatter}

There is substantial evidence that charge and spin inhomogeneities are 
present in underdoped superconducting cuprates, where charge-ordering (CO)
takes place with formation of stripe or checkerboard textures
[for a review see, e.g., Refs. \cite{review1,review2}].
On the other hand it has also been proposed that a quantum-critical instability
(quantum critical point, QCP) takes place around optimal doping 
\cite{review1,varma,tallon,CDG},
but the type of order that establishes in the low-temperature underdoped 
region is still controversial. Our proposal of spatial CO \cite{CDG}
 unifies the issue of quantum criticality with the issue
of charge inhomogeneities. According to Ref. \cite{CDG}, the CO-QCP can naturally
arise in a model where the density of
strongly interacting electrons is locally coupled to a dispersionless phonon. 
For realistic model-parameters
the CO instability first occurs along the $(1,0)$ or $(0,1)$ directions
(i.e., along the Cu-O bonds) at wavevectors of order 
$|{\bf q}_c|\approx \pi/5 \div \pi/4$. Accordingly, near the QCP the fermionic 
quasiparticles are coupled to the low-energy CO collective modes
 (CM). Strong effects occur near the ``hot'' spots at the Fermi 
surface, which are
mutually connected by ${\bf q}_c$ and reside near the $(\pm \pi, 0)$
and  $(0, \pm \pi)$ points. This is rather similar to the case 
of antiferromagnetic spin fluctuations, where, however, the critical wavevectors
are  ${\bf q}_s \approx (\pi, \pi)$. 
The phonon-driven CO instability has the additional advantage
that the phonon energy introduces a built-in energy scale, which 
provides a natural framework for non-trivial isotopic effects
\cite{andergassen,SG2}. It should be kept in mind that 
several causes like, e.g., pairing and  disorder, together with the
nearly twodimensional character of the cuprates, may prevent the formation
of static long-range CO. This gives a dynamical character to 
the excitations  accounting for the
strong variations  of the pseudogap temperature $T^*$
depending on the timescale of the probes \cite{andergassen}.
Furthermore, the charge CM  provide a ``cheap'' reservoir
of excitations, which can easily affect the various spectroscopic probes.
These effects have been extensively studied within phenomenological models,
where the fermionic quasiparticles are coupled to the CO-CM. 
In this way specific features in ARPES (e.g., the fermionic self-energy
 \cite{sulpizi}, the kink \cite{SG1} and
the isotopic dependence of the dispersions \cite{SG2}), and
in Raman \cite{CDGS}, have been coherently
explained. Remarkably, in those cases where the spectroscopic probe
provides informations on the momenta of the excitations (ARPES and
Raman), the spectra agree with the theoretical predictions only for the
CM momentum dependence corresponding to CO excitations [${\bf q}_c
\approx (\pi/2,0),(0,\pi/2)$] and {\it not} to spin excitations 
[${\bf q}_s \approx (\pi,\pi)$]. 

Recently we attacked the {\it microscopic}
calculation of the optical conductivity within a conserving
approximation.
For the clean case the diagrams correcting with the CO-CM excitations
the bare current-current response function
are reported in Fig. 1. Here we limit ourselves
to the clean case to put in evidence the effects of critical
CM's and of their typical energy scales. We use the standard 
Kubo formula to derive the conductivity directly from the
current-current response function. It is well known that this
perturbative approach without suitable diagram resummations
\cite{goetze} fails in transforming the $\delta$-like conductivity
at zero frequency into a regular Drude peak. Keeping this in mind
we concentrate on the finite-frequency conductivity, which is
adequately described by our conserving perturbative scheme.
To treat the clean case it must be considered that
the CM's are not purely electronic (otherwise the total electronic
momentum would be conserved and the conductivity at finite frequency
would identically vanish), but involve phonon degrees of freedom, which
modify the purely relaxational form of the CO-CM
$$
D({\bf q}, \omega_n)=-g^2\left[\nu ({\bf q}-{\bf q}_c)^2+|\omega_n|+
\omega_n^2/{\overline \Omega}+m\right]^{-1}
$$
where $ \omega_n$ are bosonic Matsubara frequencies, $g$ is the 
quasiparticle-CM coupling,
$\nu$ is an electronic scale, and $m\propto \xi^{-2}$ is the CO-CM mass
proportional to the inverse square correlation lenght of the CO transition.
${\overline \Omega}$ is the parameter relating electronic and phononic scales 
\cite{capDG} and it encodes the dynamical nature
of the phonon propagators. For ${\overline \Omega}<\infty$ 
 a finite dynamical response function arises, 
with various regimes for 
the optical conductivity curves.
\begin{figure}
\vspace{-0.5truecm}
\includegraphics[scale=0.25]{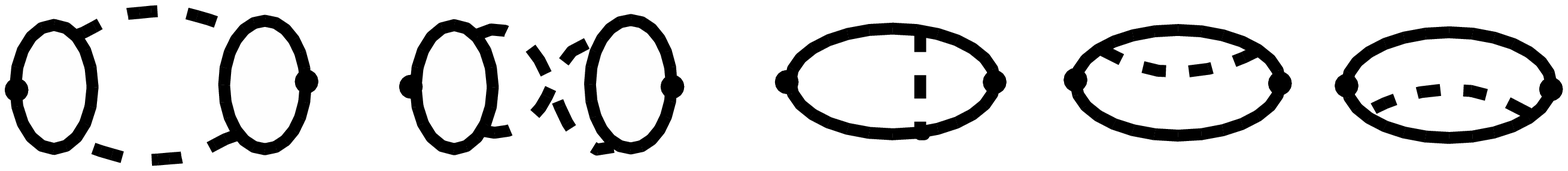}
\vspace{-4truecm}
\caption{Diagrams for the current-current response function. 
The full dots represent the current vertices, the solid and dashed
lines represent quasiparticle and CO-CM propagators respectively.
}
\label{fig.1}
%
\end{figure}
\begin{figure}
\includegraphics[scale=0.5]{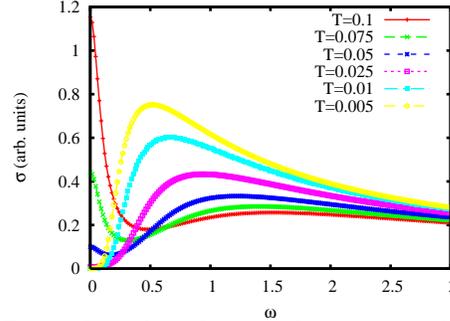}
\vspace{-0.4truecm}
\caption{Optical conductivity for a system 
of quasiparticles coupled to CO-CM modes in the quantum-critical
regime with $m=30T$. The total intensity depends on the parameter
${\overline \Omega}$ and energy units are chosen to give ${\overline \Omega}=0.1$.
}
\label{fig.2}
%
\end{figure}
Here, assuming a CM-mass linearly scaling with temperature, 
as appropriate for the quantum-critical regime,
we calculated the finite frequency contribution $\sigma(\omega)$. 
Spectral weight is generically subtracted 
from the low-frequency part of the spectra and upon decreasing $T$
peaks appear. For $m< \overline \Omega$ 
the peak saturates at $\omega \sim 
\overline \Omega$ and there is no relation between $m$ and the peak position.
On the contrary for $m>\overline \Omega$ (see Fig. 2) the peak
occurs at $\omega \sim m$ and accordingly
shifts to lower frequencies with increasing total intensity. 
This would be a distinctive signature of a quasi-critical
CM in $\sigma(\omega)$.

In summary, from our analysis one sees that, despite their neutral
character, CO-CM may strongly affect the optical conductivity. However,
 the spectra are non universal and their scaling properties only occur in
specific regimes.



\end{document}